\documentclass[fleqn,twoside,letter]{article}
\usepackage{espcrc2,natbib,graphicx,url}

\newcommand{\farcs}{\ensuremath{.\!\!^{\prime\prime}}}
\newcommand{\arcsec}{\ensuremath{^{\prime\prime}}}
\newcommand{\aap}{A\&A }
\newcommand{\mnras}{MNRAS }
\newcommand{\apjl}{ApJ }
\newcommand{\apj}{ApJ }
\newcommand{\aj}{AJ }
\newcommand{\araa}{ARA\&A }

\title{What ignites optical jets?\thanks{To appear in the proceedings
of the workshop ``Relativistic jets in the Chandra and XMM Era'' (New AR)}
\thanks{This work was supported by the
U.S. Department of Energy under contract No. DE-AC02-76CH03000.}}
\author{S.\ Jester
\address{Fermilab MS 127, PO Box 500, Batavia IL, 60510, USA ({\tt jester@fnal.gov})}}

\begin{document}
\begin{abstract}
The properties of radio galaxies and quasars with and without optical
or X-ray jets are compared.  The majority of jets from which
high-frequency emission has been detected so far (13 with optical
emission, 11 with X-rays, 13 with both) are associated with the most
powerful radio sources at any given redshift.  It is found that
optical/X-ray jet sources are more strongly beamed than the average
population of extragalactic radio sources.  This suggests that the
detection or non-detection of optical emission from jets has so far
been dominated by surface brightness selection effects, not by jet
physics.  It implies that optical jets are much more common than is
currently appreciated.
\end{abstract}



\maketitle

\section{Introduction}
\label{s:intro}

In the standard model for Active Galactic Nuclei (AGN), jets transfer
mass, energy (kinetic and electromagnetic) and momentum from the
central object into the surrounding medium.  \citet{LiuZhang02}
recently compiled a list of radio jets from the literature.  Adopting
the definition from \citet{BP84}, they list 925 jets associated with
661 radio sources.  Most of the jets have been identified by their
radio synchrotron emission. Only a few of them also show optical
synchrotron emission. Based on the most recent list of optical jets by
\citet{SU02} and the list of X-ray jets by \citet{HK02}, I have
compiled a list of 26 jets from 26 sources which have optical emission
associated with them\footnote{The full list with references is
available online at \url{http://home.fnal.gov/~jester/optjets} and
will be contained in a forthcoming publication.  The following optical
jets are not in \citet{SU02}: 3C\,31, PKS 0637$-$752, B2 0755+37, PKS
1136$-$135, PKS 1150+497, 3C\,279, 3C\,293, PKS 1354+195, 3C\,303, B2
1553+254, B2 1658+30A, 3C\,380.}.

Only a few optical jets have been studied in detail, most notably M87
\citep[e.g.,][]{MRS96,Pereta99,Per01} and, in fewer bandpasses,
3C\,273 \citep[e.g.,][ and references therein]{RM91,JesterDiss}, but
they have hardly been considered as a class of objects.  The small
number of known optical jets suggests the conclusion that they are
very rare objects.  This would, in fact, be expected from synchrotron
physics: the magnetic fields expected in jets are of order of a few
nanoTesla. This means that the loss timescales for electrons emitting
optical synchrotron radiation are of order 1000 years or shorter,
while typical arcsecond-scale jets have lengths from a few up to a few
hundred kiloparsec.  Thus, it seems natural that the frequency
$\nu_\mathrm{c}$ at which a jet's power-law spectrum cuts off lies
below the optical wavelength region in most cases.

However, in most jets where optical emission has been observed, it
extends over scales much larger than the lifetimes of ``optical''
electrons would permit.  \citet{SU02} showed that \emph{on average},
optical jets are sufficiently relativistic to remove the lifetime
discrepancy by time dilation, without the need for reacceleration of
particles within the jets \citep[before this,][ had already shown that
all optical jets known then are probably beamed]{Spaeta95}.  Still, at
least in 3C\,273's jet in-situ acceleration is necessary to explain
the observed optical synchrotron emission, in particular because very
strong beaming increases the energy loss by inverse Compton scattering
of microwave background photons \citep{Jes01}.  Moreover, the X-rays
from many jets are explained as synchrotron emission as well
\citep{HK02} -- with much shorter loss timescales than in the optical,
and correspondingly a more severe lifetime discrepancy which is harder
to remedy by relativistic effects.

It appears thus that the number of optical and X-ray jets is small
because some special physical conditions are necessary either to
switch on the extended acceleration mechanism (whose nature remains
elusive), or to accelerate the corresponding electrons near the AGN's
core and allow their escape to kiloparsec scales.  However, in direct
contradiction to this, \citet{Capeta00} noted that the detection rate
of optical jets in the B2 radio survey implies they are ``not
particularly rare''.

Here I make an attempt to find out what the required physical
conditions are, and indeed whether they are special.  Like
\citet{Spaeta95}, I find that beaming plays a role, and that more work
is necessary before we can get beyond the observer-dependent beaming
to the intrinsic physical conditions.\footnote{The cosmology is that
used by \citet{LiuZhang02}: $H_0 = 100$\,km/s/Mpc, $q_0 = 0.5$.}

\section{A case study: 3C\,273}
\label{s:3c273}

In this place, I only briefly mention an idea which could help to
explain the onset of bright high-frequency emission in the jet in
3C\,273, where beaming alone cannot explain the observed optical
emission. This jet has a divided appearance, with a continuous radio
jet extending out to 21\farcs4 projected distance from the core, but
with bright optical and X-ray emission only beyond 12\arcsec\ from the
core.  In Jester \& Krause \citetext{ApJ, in preparation}, we suggest
that the jet may be lit up by making a transition from being overdense
to underdense, with a strong shock forming near the transition region.
We show a 2.5D hydrodynamical simulation supporting this suggestion.
Full details will be contained in our forthcoming publication.

\section{Optical jets: a beaming selection effect?}
\label{s:beaming}

Synchrotron radio sources are usually assumed to have approximately
power-law spectra, with a steep high-energy cutoff at some frequency
$\nu_\mathrm{c}$ above which there is no significant emission.  An
objective definition of an ``optical'' jet would require a cutoff
frequency beyond the near-infrared.  A statistically complete survey
for optical jets would include a surface brightness limit.  Such a
survey does not exist.  I have therefore divided the radio jet list by
\citet{LiuZhang02} into subsets: the 26 jets with detected optical
emission, and the remainder as ``radio-only'' jets. This is in effect
a qualitative definition of an ``optical'' jet.

As a third class, I consider X-ray jets (this class consists entirely
of objects already included in the previous two classes). Seven radio
jets show X-ray emission, but no optical, and thirteen jets have both
optical and X-ray emission.  I am not considering optical or X-ray
emission from hot spots, only from the jet beams themselves.

\begin{figure*}[t]
\includegraphics[width=0.49\hsize]{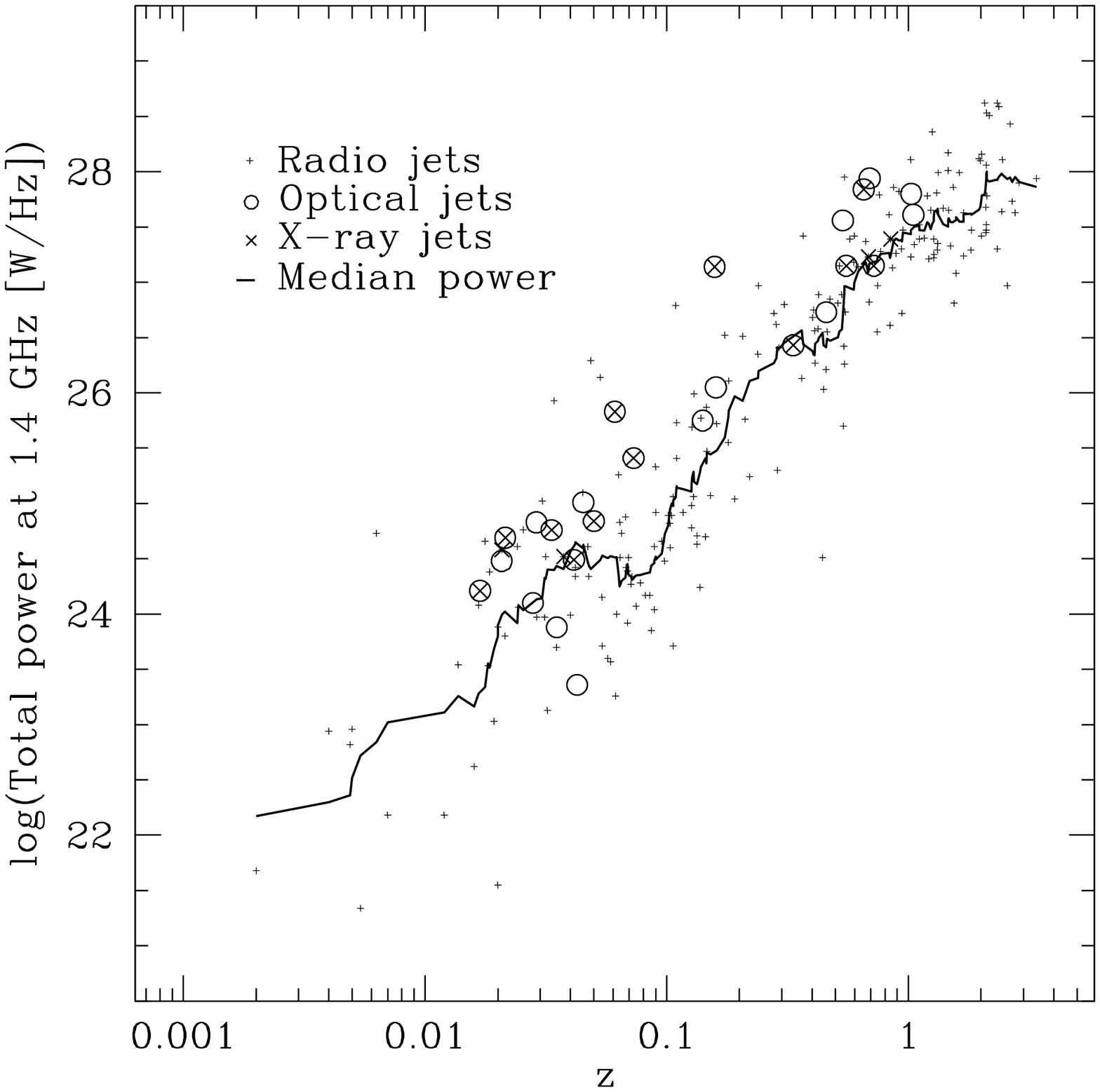}
\includegraphics[width=0.49\hsize]{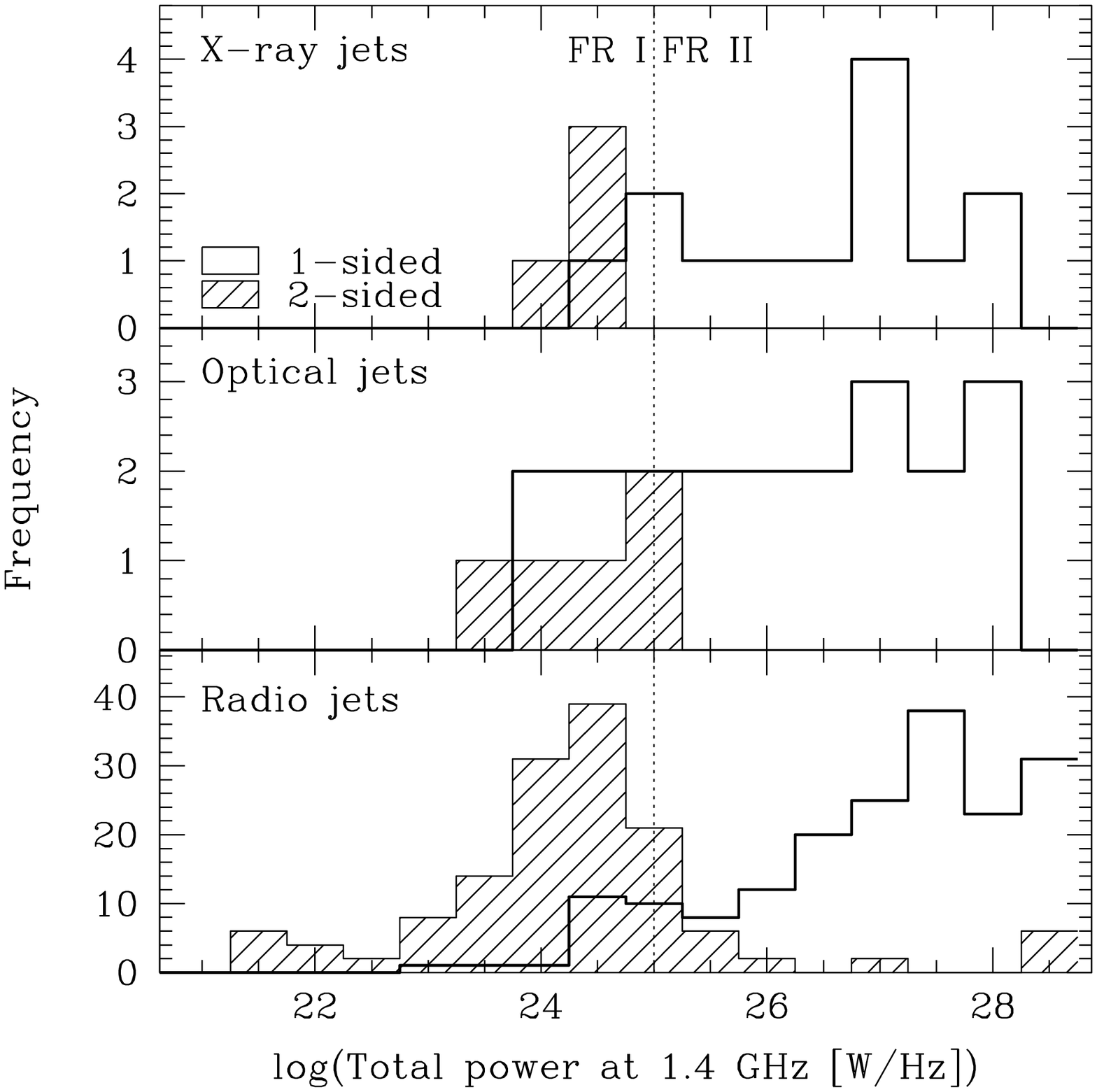}
\caption{\small {\em Left:} Core power as function of redshift for
extragalactic radio sources with jets \protect\citep[data from
][]{LiuZhang02}. The median radio power for galaxies without optical
jets (solid line) was determined at each redshift by boxcar-smoothing
the radio power over the 15 nearest neighbours in redshift. {\em
Right:} Histogram of radio powers for sources with jets, divided by
jet sidedness (``1-sided'' jets are (roughly) those in which the
jet:counterjet ratio exceeds 4, while all other jets are labeled as
``2-sided'').  The dotted line shows fiducial power separating FRI and
FRII sources.  Note that not all sources from the Hubble diagram on
the left have a sidedness determined.}
\label{f:hd}
\end{figure*}
I first consider whether the presence of optical emission from the jet
is related to the core's radio power.  Figure\,\ref{f:hd} shows that
at fixed redshift, nearly all optical jets are associated with radio
sources which are more powerful than the median at that redshift.  The
same is true for X-ray jets.  Conversely, among the sources at a fixed
radio power, only the lower-redshift sources have optical or X-ray
emission. Thus, there is no intrinsic correlation between core power
and the presence of high-frequency emission.

The histograms in Figure~\ref{f:hd} show that optical and X-ray jets
are more likely to be 1-sided than an average radio jet, although this
arises in part from the correlation between sidedness and jet power
\citep{LiuZhang02}.  Jet sidedness is of course an indicator for
relativistic beaming.

\begin{figure}[t]
\includegraphics[width=\hsize]{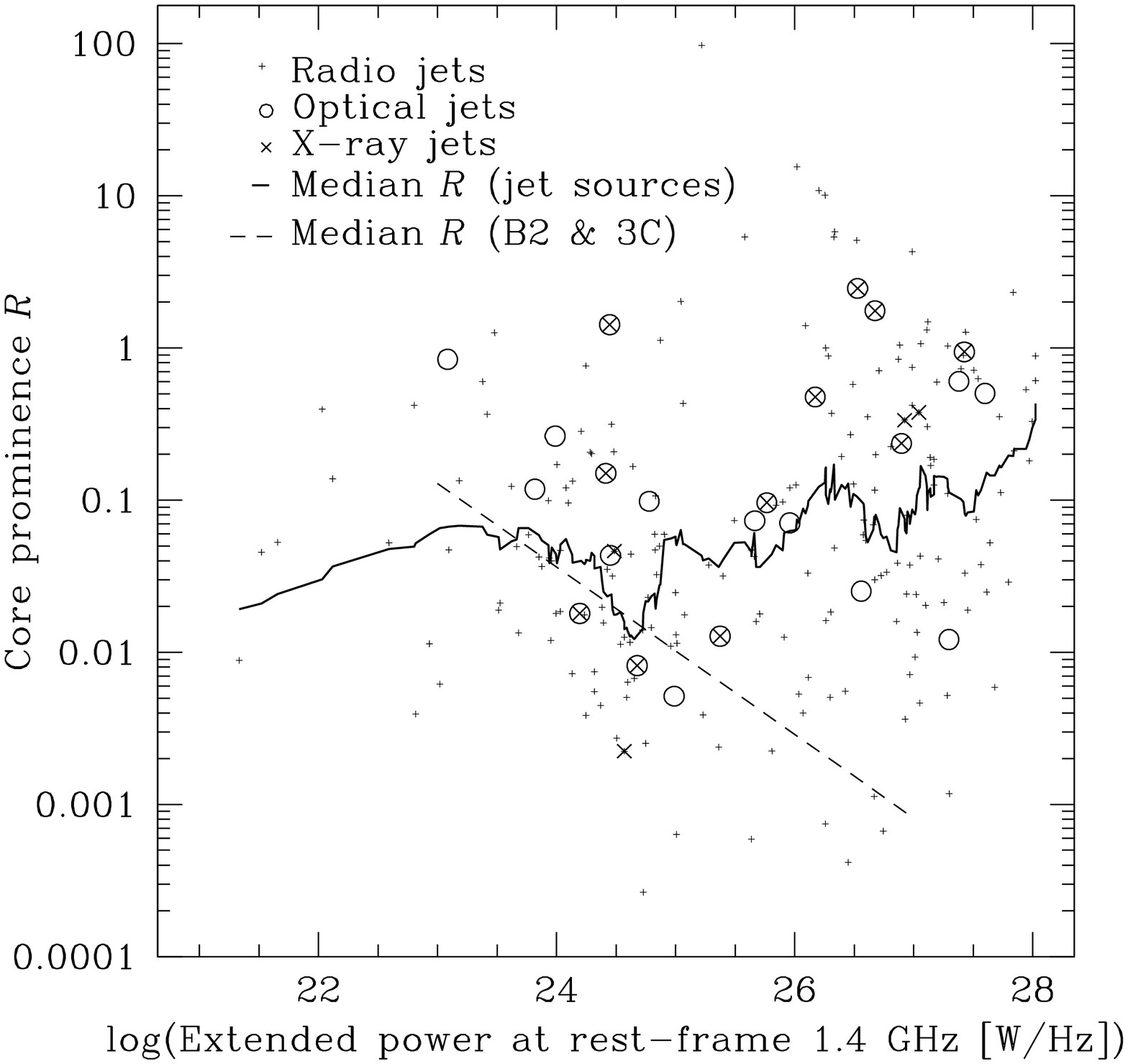}
\caption{\small Core prominence, $R$, against extended power at rest-frame
1.4\,GHz, $P_{\mathrm{e}}^{(1.4)}$. The solid line is the median core
prominence at each radio power and is obtained by boxcar-smoothing
over 19 neighbouring radio sources in extended power.  The dashed line
shows the median relation found by \protect\citet{deRuieta90} for the
low-power B2 sample together with the nearby sources from the 3C
catalogue (i.e., including both FRI and FRII sources): $\log
R_\mathrm{m} = -0.55 \log P_{\mathrm{e}}^{(1.4)} + 11.76$. It is
plotted over the range in power for which it has been determined by
\protect\citet{deRuieta90}.}
\label{f:beaming}
\end{figure}
Another beaming indicator is the core prominence ratio $R$, the ratio
of core power (affected strongly by beaming) to extended power (not
affected strongly) at a fixed rest-frame frequency
\citep[e.g.,][]{Laieta99}.  Figure~\ref{f:beaming} shows the core
prominence ratio of the three classes of jets as function of extended
power.  I have computed $R$ from the observed-frame values for core
power at 5\,GHz and total power at 1.4\,GHz listed by
\citet{LiuZhang02}, assuming a flat core spectrum and $f_\nu \propto
\nu^{-1}$ for the extended component.

\citet{SU02} have argued that the emission from optical jets is indeed
on average beamed up.  But Figures \ref{f:hd} and \ref{f:beaming}
suggest a stronger statement: First, all sources with detected radio
jets are beamed.  Secondly and more importantly, optical and X-ray
jets are \emph{more strongly} beamed than ``radio-only'' jets.  By
comparison of the jet points with the median relation for all radio
sources (dashed line in Fig.\,\ref{f:beaming}), sources with radio
jets in turn seem more strongly beamed than the radio source
population as a whole. However, in interpreting this result, one
should remember the inhomogeneity of the jet data set and the
relatively crude spectral corrections.

Since the X-ray emission of some jets is explained as beamed inverse
Compton scattering of microwave background photons \citep[see the
overview by][]{HK02}, it is expected to find strong beaming for some
of the X-ray jets. However, optical emission is usually shown or assumed to
be synchrotron emission \citep[with the possible exception of the
ultraviolet excess observed from 3C\,273's jet; ][]{Jes02}, without
reference to beaming.  While there may be a physical connection between
the jet's bulk Lorentz factor and the intrinsic optical brightness
(because more kinetic energy is available to be converted into
relativistic particles, for example), I propose here that the sample
of optical jets detected so far has overall no special physical
significance, but they have only been detected because their intrinsic
surface brightness has been beamed up sufficiently.

Just before the time of this writing, a paper appeared as preprint
which shows that exactly this surface brightness selection effect can
account for the optical non-detection of all but two jets in HST
imaging of 57 sources from the B2 sample \citep{Pareta02}.  If this
holds for all other searches, it implies that optical jets are much
more common in all extragalactic radio sources than has been
appreciated so far, and they have not been detected in significant
numbers simply because they have not been looked for with sufficient
surface brightness sensitivity.

\section{Conclusions: Finding more optical jets}
\label{s:conc}

The present comparison of the radio properties of sources with
detected optical or X-ray jets to those with ``radio-only'' jets has
not brought to light any connection between intrinsic physical
properties and the presence of optical or X-ray emission.  Instead, it
is found that optical jets are more strongly beamed on average than
other radio jets are.  I therefore suggest that the entire sample of
presently known optical jets is determined mainly by surface
brightness selection effects.  Like \citet{Pareta02} have done for the
B2 sample, this suggestion has to be tested by comparing the optical
surface brightness expected from an extrapolation of radio maps with a
suitable power law ($f_\nu \propto \nu^{-0.7}$ represents most jet
spectra well) to the surface brightness limits of available optical
observations.  For those jets without suitably deep observations, the
most efficient search method will employ observations in the
near-infrared. In this way, jets with cutoffs just below the optical
are still detectable.  Positive detections will be followed up by
observations in a second bandpass to determine the spectral index, and
hence the detectability at higher frequencies.  Only with a firm
sample of detections \emph{and} non-detections can we analyse the
dependence of the jet's synchrotron cutoff frequency $\nu_\mathrm{c}$
on other observables and thus address the question of what ignites
optical jets.

\hspace{4ex} I am grateful to Martin Hardcastle and Robert Laing for
valuable comments.


\begin{thebibliography}{16}
\expandafter\ifx\csname natexlab\endcsname\relax\def\natexlab#1{#1}\fi
{\small
\bibitem[{{Bridle} and {Perley}(1984)}]{BP84}
{Bridle}, A.~H., {Perley}, R.~A., 1984. \araa 22, 319.
\bibitem[{{Capetti} et~al.(2000){Capetti}, {de Ruiter}, {Fanti}, {Morganti},
  {Parma}, and {Ulrich}}]{Capeta00}
{Capetti}, A., et~al., 2000. \aap 362, 871.

\bibitem[{{de Ruiter} et~al.(1990){de Ruiter}, {Parma}, {Fanti}, and
  {Fanti}}]{deRuieta90}
{de Ruiter}, H.~R., et~al. 1990. \aap 227,
  351.

\bibitem[{{Harris} and {Krawczynski}(2002)}]{HK02}
{Harris}, D., {Krawczynski}, H., 2002. \apj 565, 244.

\bibitem[{{Jester}(2001)}]{JesterDiss}
{Jester}, S., 2001. Ph.D. thesis, U. Heidelberg,
  \url{http://www.ub.uni-heidelberg.de/archiv/1806}.

\bibitem[{{Jester} et~al.(2002){Jester}, {R\"oser}, {Meisenheimer}, and
  {Perley}}]{Jes02}
{Jester}, S., et~al., 2002. \aap
  385, L27.

\bibitem[{{Jester} et~al.(2001){Jester}, {R\"oser}, {Meisenheimer}, {Perley},
  and {Conway}}]{Jes01}
{Jester}, S., et~al., 2001. \aap 373, 447.

\bibitem[{{Laing} et~al.(1999){Laing}, {Parma}, {de Ruiter}, and
  {Fanti}}]{Laieta99}
{Laing}, R.~A., et~al. 1999. \mnras
  306, 513.

\bibitem[{{Liu} and {Zhang}(2002)}]{LiuZhang02}
{Liu}, F.~K., {Zhang}, Y.~H., 2002. \aap 381, 757.

\bibitem[{{Meisenheimer} et~al.(1996){Meisenheimer}, {R\"oser}, and
  {Schl\"otelburg}}]{MRS96}
{Meisenheimer}, K., {R\"oser}, H.-J., {Schl\"otelburg}, M., 1996. \aap
  307, 61.

\bibitem[{{Parma} et~al.(2003){Parma}, {de Ruiter}, {Capetti}, {Fanti},
  {Morganti}, {Bondi}, {Laing}, and {Canvin}}]{Pareta02}
{Parma}, P., et~al., 2003. \aap 397, 27 
  (astro-ph/0210461).

\bibitem[{{Perlman} et~al.(2001){Perlman}, {Biretta}, {Sparks}, {Macchetto},
  and Leahy}]{Per01}
{Perlman}, E.~S., et~al., 2001. \apj 551, 206.

\bibitem[{{Perlman} et~al.(1999){Perlman}, {Biretta}, {Zhou}, {Sparks}, and
  {Macchetto}}]{Pereta99}
{Perlman}, E.~S., et~al., 1999. \aj 117, 2185.

\bibitem[{{R\"oser} and {Meisenheimer}(1991)}]{RM91}
{R\"oser}, H.-J., {Meisenheimer}, K., 1991. \aap 252, 458.

\bibitem[{{Scarpa} and {Urry}(2002)}]{SU02}
{Scarpa}, R., {Urry}, C.~M., 2002. New Astronomy Review 46, 405.

\bibitem[{{Sparks} et~al.(1995){Sparks}, {Golombek}, {Baum}, {Biretta}, {de
  Koff}, {Macchetto}, {McCarthy}, and {Miley}}]{Spaeta95}
{Sparks}, W.~B., et~al., 1995. \apjl 450, L55.
}
\end{thebibliography}
\setlength{\bibsep}{0ex}

\end{document}